# Judgment in macroeconomic output growth predictions
# Efficiency, accuracy and persistence

Michael Pedersen

Adolfo Ibáñez University, Business School

Viña del Mar, Chile

michael.pedersen@uai.cl

**Abstract**

The present study applies observations of individual predictions of the first three releases of the US output growth rate to evaluate how the applied judgment affects prediction efficiency and accuracy as well as if judgment is persistent. While the first two issues have been assessed in other studies, there is little evidence on the formation of judgment in macroeconomic projections. Most of the forecasters produce unbiased predictions, but employing the median Bloomberg projection as baseline, it turns out that judgment generally does not improve accuracy. There seems to be persistence in the judgment applied by forecasters in the sense that the sign of the adjustment in the first release prediction carries over to the projections of the two following revisions. One possible explanation is that forecasters use some kind of anchor-and-adjustment heuristic.

# 1. Introduction

When making predictions of macroeconomic variables, professional forecasters, from the private as well as the public sector, typically employ macroeconomic and/or econometric models, which they load with all relevant information. The results of the different models are then analyzed by the forecaster, who applies judgment to come up with the final forecast.[1] This judgment is subjective and probably affected by several economic and non-economic[2] factors, of which there is little knowledge since it is not possible to observe which part of the forecast is model-based and which part is judgment. The literature is, however, growing and the present study contributes to the research by analyzing the judgment included in backcasts[3] of the US output growth rate made by economists who report their predictions to Bloomberg. The results suggest that the majority of the analysts make unbiased predictions, but the judgment only improves projection accuracy for less than one third of the economists. One reason may be that judgment seems to be persistent, i.e. forecasters whose judgments are biased in a particular direction tend to include the same bias in the next prediction.

The benefits of augmenting mechanical forecasts with judgment have been assessed by researchers in several contexts and disciplines.[4] A challenge is that judgment, generally, is private knowledge and unobservable for externals to the forecasting process. So are the models employed for obtaining the baseline prediction, which guide the final projection. Hence, some assumptions have to be adopted with respect to the baseline forecast, which will be common for all forecasters. Admittedly, it is a strong assumption that all forecasters should employ the same baseline projection and because of lack of information with respect to the individual forecast processes, Galvão et al. (2021) relies on statistical models for providing the baseline prediction.

---

[1] Concerning projections made by private companies, Fildes & Petropoulos (2015) find that the majority make forecasts using a combination of statistical models and judgmental methods. Pedersen (2020) surveys financial traders about how they form their inflation forecasts. The majority employ several methods, which suggests an important judgmental component in the final prediction.

[2] In a study of anomalies in inflation and output growth prediction errors, Pedersen (2019) finds that the median projection is not efficient and that the inefficiency can be explained by day effects and general optimism / pessimism captured by current consumer and firm sentiments.

[3] The term "backcast" refers to the prediction of a variable (the output growth rate in the present context) for a period that has already passed. The value of the variable has, however, not been published at the time of making the prediction.

[4] International Journal of Forecasting has dedicated three special sections to the subject. See the editorials in Parackal et al. (2007), Fildes & Goodwin (2013) and Bolger & Wright (2017).

The present study employs the median Bloomberg projection as the baseline point backcast. There is of course a caveat when using a statistical measure, such as the median, since it is calculated with final predictions, i.e. those that include judgment. The implication of utilizing the median backcast is that up to half of the judgments are negative and up to half of them are positive. If forecasters use all available information to come up with a mechanical prediction that is optimal in terms of accuracy, the assumption that the judgments follow an almost symmetrical distribution does not seem to be too restrictive. Combining several forecasts have been argued to improve accuracy[5] and, hence, it seems appropriate to employ the median as an optimal prediction, i.e. the common baseline projection upon which the economists apply individual judgment. This is then defined as the difference between the individual forecast and the median.

The objective of the paper in hand is twofold. The first contribution relates to the literature that analyzes the extent to which judgment improve macroeconomic forecast accuracy. While this literature is currently evolving, it is still scarce. In some early studies on the subject, Turner (1990) argues that judgment has large effect on forecasts, while McNees (1990) finds that judgment generally improve accuracy, but there is a tendency to over-adjust. Clements (1995) and Fildes & Stekler (2002) also discover that adjustment of model-produced forecasts tends to improve accuracy, but Clements finds no impact on forecast rationality. Galvão et al. (2021) analyze the effect of judgment on density forecasts and argue that applied to point forecasts, it tends to improve density forecast accuracy at the short horizon and in times with elevated uncertainty.

Rather than considering forecasts of future events, the present application studies the effect of judgment on backcasts of the output growth rate for the previous quarter. Hence, there is no uncertainty with respect to shocks that may affect updates of the predictions, but there may be surprises in data published for the previous period. Predictions are evaluated for the first growth estimate published by the Bureau of Economic Analysis (BEA) as well as for the first two revisions. The second contribution of this paper is an assessment of the extent to which persistence is present in the judgment applied by the forecasters. Even though research has acknowledged that judgment is an important component of macroeconomic forecasts, little, if any, attention has been given to the formation of judgment, for example if forecasters learn from previous mistakes or if judgment is persistent. Evaluating judgment applied to backcasts of three releases of the same output growth rate comprises a good framework for analyzing the formation process.

---

[5] See Wang et al. (2023) for a recent review of the literature on forecast combinations. Önkal et al. (2019) find in a survey with business executives that combination of forecasts is a factor that enhances trust in the projection.

While there is little knowledge of the formation of judgment in macroeconomics forecasts, experiments and analyses from other disciplines provide some insight into why forecasters augment model-predictions with judgment and which factors influence the way the judgment is made. In general, people seem to prefer predictions made by humans rather than computers since they understand better the reasoning of person-made predictions and seem to have larger tolerance for errors made by humans.[6] Önkal & Gönül (2005) argue that forecasters adjust mechanical projections for egocentrical reasons, claiming ownership of the prediction and justify their role in the process. Model-based forecasts are designed to capture and process long-term tendencies in data, while humans are capable of observing recent shocks and evaluate the impact on the target variable. This justifies adjustments to mechanical forecasts.[7] In this context, forecasters may apply judgment heuristics or more specifically anchor-and-adjustment heuristic, which some evidence suggests that people who apply time-series for forecasting use.[8] The extent to which this also applies in macroeconomic forecasting still has to be discovered, but the present study takes a first step in that direction by analyzing if persistence is part of the forecasters' judgment formation processes.

The outline of the remainder of the paper is as follows: Section 2 details the data employed in the analysis and presents some descriptive statistics. Section 3 explains in details how judgment is measured and discusses if the baseline model provides unbiased and efficient predictions. Based on these observations, it continues with an assessment of the extent to which judgment improve prediction accuracy. Section 4 addresses how judgment may be persistent over time by means of unbalanced panel regressions. Section 5 concludes.

---

[6] See Bonaccio & Dalal (2006), Dietvorst et al. (2015), and Önkal et al. (2009).
[7] See Armstrong & Collopy (1998) and Goodwin (2002) concerning organizational forecasts. According to the findings of Bolger & Harvey (1993) forecasters tend, however, to put too much weight on such events.
[8] See Hogart & Makridakis (1981) and Lawrence & O'Connor (1992).

## 2. Data and descriptive statistics

The data consist of predictions of the US real output growth rate (annualized quarter-over-quarter) from the Bloomberg database. Backcasts of the first, second and third BEA estimates are collected from the first quarter 2000 to the fourth quarter of 2022.[9] After cleaning data,[10] the database contains 6,827 first estimate predictions, 6,349 second estimate projections and 6,151 backcasts for the third release. The observations are distributed among 267 firms and 283 economists. Table 1 presents some descriptive statistics.

[Table 1]

In general, there are more predictions of the first growth estimate than of the following two and the forecast error, measured by the Average Root Mean Squared Error (ARMSE),[11] is typically larger, which is as expected since more information is available the later the backcast is made. Indeed, there is less variation in the projections of the later releases, i.e. less uncertainty. While there is no indication of general skewness in the quarterly distributions of the projections, there seems to be periods in which the backcast distribution is skewed to the left as well as to the right. Even though there is more variation in the first estimate predictions, the average kurtosis is thinner than those of a normal distribution. The average kurtoses are higher for the projections of the second and third estimates.

Fig. 1 elaborates further on the moments of the distributions by plotting the second to the fourth moments across time. Indeed, there are some spikes in the variances, i.e. differences in the disagreement amongst forecasters, across time. When applying a common baseline projection, strong disagreement implies extensive use of different judgments in the backcasts processes. In general, projections do not seem to be positively or negatively biased, but there are some quarters with extreme projections as illustrated by the spikes in the kurtosis. To sum up, the three moments shown in the illustrations in Fig. 1 indicate that Bloomberg forecasters apply judgment actively when projecting different estimates of the US growth rates.

---

[9] First release backcasts for the third quarter of 2000 and second release predictions from the fourth quarter of 2018 were not available.

[10] The cleaning process consisted mainly in deleting observations registered for the same economist at the same quarter or re-assigning different firm id for observations for the same firm the same quarter, but with different economists. Very few observations where either altered or deleted. Details are available upon request.

[11] $ARMSE^k = \frac{1}{T_k}\sum_{t=1}^{T_k} RMSE_t^k$, where $T_k$ is the number of quarters where predictions of release $k$ are available and $RMSE_t^k = \sqrt{\frac{1}{N_t}\sum_{i=1}^{N_t}(E_i(y_t^k) - y_t^k)^2}$, where $E_i(y_t^k)$ is the backcast of the $k$´th release of $y_t$ made by economist $i$ and $N_t$ is the number of projections available at time $t$.

[Figure 1]

Fig. 2 illustrates the prediction errors, represented by the RMSE of the forecasters' projections during the period analyzed. Typically, the first estimate errors are substantially larger than the those of the other estimates, but there are a few episodes where this is not the case.

[Figure 2]

Some predictions are only contributed to a company and not a specific economist or group of economists. In the context of evaluating subjective judgment, it is preferred to work with the backcasts to which there are assigned specific economists assuming that they are the main responsible for the projections and, hence, the applied judgment. As reported in Table 2, the final database consists of 6,307 predictions of the first BEA estimate made by 272 economists (or groups of economists), which is more than for the other two estimates. The numbers for joint projections, i.e. of more than one estimates for the same quarterly growth rate, are also presented in the table.

[Table 2]

Not all the economists report predictions every quarter and there is rotation in the panel of forecasters. As shown in the three last rows of Table 2, approximately 50 forecasters have supplied predictions more than 50% of the quarters in the sample. If the requirement is only 25% of the periods, the number of participants increases to around 100.

# 3. Judgment

A typical forecasting process consists in feeding econometric models with data and then evaluating and combining the projections of these models. The final outcome of this mechanical process will be referred to as the baseline prediction. In order to analyze the judgment of several forecasters it is necessary to define a common baseline prediction. With this in hand, the judgment of the individual forecast is defined as the deviation from the baseline.

## 3.1. A common baseline prediction

While econometric models are simple and subjective approximations of reality, they do serve as systematic tools to gather past regularities in order to predict the future. But since the models are assumption-based, they do not match the economic reality exactly and frequently forecasters employ several models based on different assumptions and then chose the combination, which they believe supplies the best outcome. The selection may be automatic, based on certain criteria, or made by judgment.[12] The professional forecaster adjusts the mechanically generated prediction based on expert knowledge, i.e. makes a judgmental adjustment.

To analyze the judgment, one must have information of the mechanical projections, which is rarely available.[13] When working with a panel of forecasters, it is necessary to define a model-based, or baseline, projection, which is representative for all of them. In the present application the median projection is chosen as a proxy for the baseline. A large share of the forecasters in the panel are employed by companies that publish projections in other media and thereby visualize their outlook. Thus, with the information available to the economists it is possible to get an estimate of the median prediction. An alternative would be to use the mean, which is also available at Bloomberg, but the median is preferred as it is less sensitive to extreme observations. Both measurements are, however, in line with the large literature that advocates for combining forecasts.[14]

---

[12] Petropoulos et al. (2022) note that judgmental selection is much more common in practice than automatic selection, which has been verified in an experiment by Önkal et al. (2009).
[13] McNees (1990) and Clements (1995) did, however, gain access to unpublished material with which they were able to evaluate the role of the judgment in macroeconomic forecasts in the US and UK.

It is admittedly a strong assumption that all forecasters employ the same model-based prediction and, as a consequence, deviations from this projection are exclusively due to individual judgment. It is likely that idiosyncratic prediction errors could be attributable to impacts (e.g. omitted variables or incorrect functional forms of the models applied) that are most likely not constant over time. But since the models and judgments applied by the forecasters are private knowledge, it is not possible to disentangle them from each other. Thus, it is necessary to define a common baseline projection that can be used to approximate the average judgment.

Professional forecasters, such as those reporting to Bloomberg, are likely to take advantages of combining the outputs of several models to produce a projection, which they can then adjust by judgment if they wish. The median employed in present study is certainly a combination of different models, but also of different judgments used to augment the model-based projections. Thus, it should be evaluated to what extent it serves as an approximation of what would be a reasonable model-based projection employed by the predicters. When developing a statistical model, or set of models, for forecasting, a desirable characteristic is that the outcome is not systematically biased with respect to the vision of the forecaster. In other words, when considering a longer period of time, one may expect that the predictor applies negative and positive judgments to obtain the final projections in more or less equal shares.

Table 3 presents some statistics of the implicit judgments applied by the individual forecasters under the assumption that the median projection is the model-based prediction for all of them. The average predictor´s assumed judgment-augmented forecast is the same as the median in 13 to 14% of the quarters for the first BEA estimate, a number which is more than twice the size for the second estimate and for the third estimate, more than half of the predictions are equal to the median. These are the situations in which the forecasters are assumed not to apply judgment, or rather that the judgment is neutral with respect to the model-based projection. Since it is the median prediction that is assumed to be the model-based one, the shares of the implicit judgments that are negative and positive, respectively, are almost the same. The standard deviations suggest, however, that there is some heterogeneity among forecasters.

---

[14] Wang et al. (2023) report that there is no consensus of whether the mean or the median is better in terms of forecast performance.

[Table 3]

To gain further insight into the implicit judgments, Fig. 3 illustrates shares of economists who implicitly have applied negative judgments in certain parts, e.g. between 20 and 40%, of their predictions. The histograms disregard episodes with neutral judgment such that the distributions of the positive judgments can be derived by mirroring the distributions of the negative. The histograms are almost symmetric and it appears that, for the first and second BEA estimates of the output growth, the backcasts of more than half of the economists are below the median in 40 to 60% of the quarters, where it is not equal to the median, and, hence, also above the median in 40 to 60% of the cases. For the third estimate, where more than half of the predictions are equal to the median, it is about one third of the them that are within the 40 to 60% interval. Hence, for most forecasters the median prediction does not appear to contain systematic biases with respect to the individual predictions and it may serve as a projection from which an approximate judgment component of individual projections can be extracted. The caveats mentioned earlier should, however, be remembered when interpreting the results presented in this study. The next subsection discusses some properties of the median and mean predictions for the period considered.

[Figure 3]

### 3.2. The median and mean projections

The test of whether the two predictions are efficient is based on the concept of rational expectations put forward by Muth (1961), i.e. a forecast is rational if its expected value is equal to the value of the variable projected, when incorporating all available information. Let, as before, $y_t^k$ denote the $k$´th BEA release of the output growth rate for quarter $t$ and $\tilde{y}_t^k$ the prediction of the $k$´th release, the median or mean of the forecasters' backcasts in the present application. Then

$$\tilde{y}_t^k = E[y_t^k|\Omega_t], \qquad (1)$$

where $\Omega_t$ is the information set at time *t*. Following e.g. Swanson and van Dijk (2006), the relation (1) can be analyzed by the regression[15]

$$y_t^k = \alpha^k + \eta^k \tilde{y}_t^k + \boldsymbol{W}_t^{k\prime} \boldsymbol{\gamma}^k + \varepsilon_t^k, \tag{2}$$

where $\boldsymbol{W}_t^k$ is vector of variables representing the available information that could have been employed to make a better prediction, i.e. omitted variables. Greek letters denote coefficients to be estimated and $\varepsilon_t^k$ is an error term that is assumed to be uncorrelated with $\tilde{y}_t^k$ and $\boldsymbol{W}_t^k$. As in Swanson and van Dijk, this study applies the regression (2) formulated in prediction errors:

$$y_t^k - \tilde{y}_t^k = \alpha^k + \beta^k \tilde{y}_t^k + \boldsymbol{W}_t^{k\prime} \boldsymbol{\gamma}^k + \varepsilon_t^k, \tag{3}$$

where $\beta^k = \eta^k - 1$. If assuming $\boldsymbol{\gamma}^k = \boldsymbol{0}$, the hypothesis of unbiasedness is formulated as $\alpha^k = \beta^k = 0$, while the more general hypothesis of efficiency is $\alpha^k = \beta^k = 0$ and $\boldsymbol{\gamma}^k = \boldsymbol{0}$.

In the present context, where the variables included in the models of the forecasters are not known, it is not obvious which ones to include in $\boldsymbol{W}_t^k$. This is particularly a concern when considering ensemble predictions, like the median and the mean, that are computed from a number of inputs. Two projections, which are easily accessible at the time of producing the prediction, are the latest nowcast of the Survey of Professional Forecasters (SPF) and forecasts made with a simple autoregressive (AR) model. It should be mentioned, however, that several of the Bloomberg forecasters do also supply projections to the SPF, but they are nowcasts delivered months before the backcasts reported to Bloomberg. But since both projections are known to the forecasters, they should not contain extra information useful for the predictions, examined in the present study, if they are efficient.

The data of the SPF are extracted from the webpage of the Federal Reserve Bank of Philadelphia,[16] and the median (mean) nowcast is included in the efficiency test of the median (mean) Bloomberg backcast. The AR models, which include one lag[17] and an intercept, are estimated recursively with observations of the three BEA releases from the third quarter 1965.[18] Table 4 presents the results as well as some measures of forecast accuracy.

---

[15] The regression (2) has been applied in several applications in the literature on data revisions, such as those of Mankiw et al. (1984) and Keane & Runkle (1990).

[16] https://www.philadelphiafed.org/surveys-and-data/real-time-data-research/survey-of-professional-forecasters.

[17] One lag is supported by the Akaike, Schwarz and Hannan-Quinn information criteria for all releases in all periods, except for the last nine where all criteria indicated models with no lags.

[Table 4]

Generally, the results for the median and the mean are similar. While both predictions appear to be unbiased, there is no strong evidence of efficiency in the sense that information contained in the SPF and / or the AR projections have statistically significantly impact on the backcast errors. A more granular analysis shows that both projections impact the Bloomberg predictions of the first BEA estimate, while only the SPF nowcast affect those of the second estimate and only the AR forecasts explain part of the prediction errors of the third release. Also, with respect to prediction accuracy the median and the mean are practically identical.[19]

### 3.3. Do judgments improve projections?

With the baseline prediction in hand, it is now possible to evaluate the judgment of the forecasters. This is done by testing backcast unbiasedness and efficiency as well as evaluating accuracy of the predictions of the individual forecasters.

It is indeed possible to improve upon the median projection, i.e. potentially the judgment augmented prediction can be more accurate than the baseline. The median is only correct in 3, 18 and 26% of the quarters for the first, second and third estimate, respectively. With respect to the first BEA release, the median tends to overpredict (56% of the cases), while it slightly underpredicts the second (44%) and the third (39%) estimates.[20]

Table 5 shows how many of the economists have unbiased and efficient backcasts when employing the regression (3) and testing the hypothesis $(\alpha^k, \beta^k) = (0,0)$ for each of them for unbiasedness and $(\alpha^k, \beta^k) = (0,0)$ and $\boldsymbol{\gamma^k = 0}$ for efficiency. For this exercise the sample is restricted to include only those who have reported expectations in more than 10% of the quarters. It cannot be rejected that approximately three quarters of the analysts make unbiased predictions of the two first BEA releases. For the third estimate it is 88% of the economists that produce unbiased backcasts. The percentages diminish for the last two releases when restricting the sample to those with more predictions, but even so, there is no strong evidence to indicate that the application of judgment generally damages the predictions with respect to bias.

---

[18] Data are extracted from the webpage of the Federal Reserve Bank of Philadelphia (https://www.philadelphiafed.org/surveys-and-data/real-time-data-research/first-second-third). Three missing observations were calculated with extra- and interpolation.
[19] This was also supported by the accuracy comparison tests of Diebold & Mariano (1995) and Harvey et al. (1997).
[20] This is in contrast to evidence found in the organizational forecast literature. Fildes et al. (2009) report that adjustments in a negative direction in four large supply-chain companies improved accuracy.

[Table 5]

Concerning efficiency, it is approximately half of the economists that construct efficient predictions, which implies that the judgment applied with respect to the median prediction has helped in the respect. When restricting the sample to those who have supplied at least one quarter of the projections, the shares descend markedly for the second and third BEA estimates. Less than one fifth of the predictions of the analysts who have reported to Bloomberg at least half of the periods in the sample are efficient.

To sum up, for the majority of the forecasters, the application of judgment does not seem to damage the unbiasedness of the baseline prediction, but this finding does not hold for those who have delivered at least half of the predictions of the second and third BEA estimates. Approximately half of the economists have been able to produce more efficient predictions than the baseline, but this result hinges on those with less backcasts.

Concerning forecast accuracy, measured by the RMSE,[21] only 19% of the economists make better predictions than the median of the first output growth estimate, numbers which are 12 and 24% for the second and third estimates. If we consider only the participants who have supplied predictions more than 10% of the quarters, the records change to 7 and 26% for the second and third estimates, while the 19% of the first release remain the same. In other words, more than 70% of the forecasters would have made more accurate projections if they had used the median as their backcast, i.e. for the main part of the analysts judgment did not improve the accuracy, which is in contrast to the findings of McNees (1990), Clements (1995), Fildes & Stekler (2002), and Galvão et al. (2021). One possible explanation for these results may be that the common baseline projection in the present study is quite accurate itself.

---

[21] Comparisons are made only with the quarters where the economists have provided predictions.

## 4. Judgment persistence

After evaluating the judgment of the forecasters in the last section, this one characterizes it in terms of relationship with that previously applied. This is done by means of non-balanced panel estimations of the type

$$j_{i,t}^k = \alpha_i^k + \beta^k X_{i,t}^k + \delta_t^k + \varepsilon_{i,t}^k, \qquad (4)$$

where $j_{i,t}^k = E_i(y_t^k) - \tilde{y}_t^k$ is the judgment applied by forecaster $i$ for the $k$´th estimate of the output growth rate of quarter $t$, $\alpha_i^k$ and $\delta_t^k$ are individual fixed effects and time effects, respectively, while $X_{i,t}^k$ is one of two different variables to characterize the persistence, the past judgment for the same release, $Lj_{i,t}^k$ ($L$ denotes the lag operator), or judgement applied in the prediction of the previous BEA estimate, $j_{i,t}^{k-1}$. In (4), $\beta^k$ is a coefficient to be estimated and $\varepsilon_{i,t}^k$ are the errors.

Alternatively, one could include past backcast errors as an explanatory variable in (4). But since the prediction error is $e_{i,t}^k = E_i(y_t^k) - y_t^k$, then the difference between the judgment and the error is $j_{i,t}^k - e_{i,t}^k = y_t^k - \tilde{y}_t^k$, which is captured by including time fixed effects in the regression.

Results of estimations of variations of (4) for the first BEA estimate are displayed in Table 6, for the second estimate in Table 7 and for the third in Table 8. With respect to the first one, persistence seems indeed to be present in the forecasters' predictions in the sense that the sign of the judgment included in the prediction of the first BEA estimate appears to be the same as the one of the same projection the previous quarter. The estimated coefficients are statistically significantly positive in all specifications and with similar sizes, which indicates that about 10% of the judgment is carried over to the next quarter projection. The results do practically not change when controlling for fixed effects related to individual forecasters nor when including fixed time dummies. On the other hand, there is no evidence that judgment included in the most recent projection, i.e. that of the third BEA estimate made the previous quarter, affects the judgment included in the first BEA release prediction.

[Table 6]

[Table 7]

[Table 8]

The results for the second and third releases change as the judgments applied for same predictions the previous quarter do not affect current projections, but so does the judgment of the prediction of the previous release for the same quarter. The positive coefficients imply that the economists tend to preserve the sign of the previous judgment, but the size of the judgment seems to be lower. This is in line with the fact that more data are available, i.e. have been published, when predictions of later estimates are made. This is also apparent in the lower variances in second and third estimate backcasts.

Summing up, the estimations suggest that forecasters' applied judgment is persistent as a negative or positive adjustment tends to carry over to the next projection. The first prediction of the output growth rate of a given quarter appears to be affected by the judgment applied in the same backcast the previous quarter, while the judgment of the second estimate prediction is influenced by that of the first and the third projection by that of the second.

The fact that forecasters' judgment is persistent could suggest that they apply some kind of heuristic judgment, which seems to be anchored in their minds.[22] This may play out in different ways.[23] If the economist, for example, has a habit of employing time series and has observed a recent trend in the BEA data revisions, then s(he) may anchor at the last point and adjust to stay at the recently observed tendency. Another example could be that the observed revisions contain no trend, but exhibit autocorrelation. The forecaster may then anchor at the last point and adjust towards the mean. These are merely a couple of possible explanations, but there may be others which can also explain judgment persistence such as habitual thinking patterns, emotional influences, which may persist over time, and overconfidence in own ability with respect to the judgment. Future research and experiments may disclose which of these factors, or combinations of them, are important for explaining judgment related to macroeconomic forecasts. In general, more investigation is needed to reveal which factors can explain the judgment, since the relatively small $R^2$ of the regressions indicate that persistence is not the only one.

---

[22] In a study of profit forecasting, Nardi et al. (2022) find that among cognitive biases, optimism is negatively related with accuracy of the projections, while anchoring bias has a positive relationship.
[23] See also section 2.11 in Petropoulos et al. (2022).

## 5. Concluding remarks

The present study applied observations of projections from Bloomberg of the first three releases of the US output growth rate for the previous quarter, covering a period of 23 years. Judgment of the predictions were analyzed with respect to the capacity of improving the backcasts as well as a particular feature of the judgment formation process, namely possible persistence. Since there exists no information of the forecasters' models, the common baseline prediction was defined as the median projection, which admittedly has some drawbacks, e.g. being determined endogenously by the projections including the judgment components, but also some strengths, such as being an objective combination of several "models". An analysis of the median prediction indicated that it does not have systematic biases with respect to the final predictions of most forecasters and it is an unbiased prediction for all three releases, but not efficient.

The large majority of the individual economists make unbiased backcasts and about half of them also makes efficient ones. But the more projections a forecaster supplies to Bloomberg, the smaller is the share of them for which the hypothesis of efficiency cannot be rejected. Judgment does not seem to improve the accuracy of the predictions. In fact, more than 70% of the forecasters would have made more precise predictions without augmenting the baseline with judgment.

As a first step to obtain more knowledge of the judgment formation process, an analysis of possible persistence in the judgment was conducted. It turned out that judgment is indeed persistent in the sense that the sign of the adjustment made for the first release carries over to the predictions of the second and third releases. This could indicate that forecasters use anchor-and-adjustment heuristics, but other explanations are also plausible.

The fact the forecast judgment appears to be persistent could have implications for the accuracy of the predictions. Whether or not this is the case depends on the underlying reason for the persistence. Some studies, e.g. Fuhrer (2018), find that survey respondents do not revise forecasts efficiently as they react insufficiently to new information. It is not known to what extent this underreaction is due to model-based forecasts or judgment, but forecasters seem to form their expectations with respect some anchor, maybe the unconditional mean, and produce the forecasts in that context. To be able to improve the forecast performances of private and public entities, clearly more analysis of the judgment process is required.

The paper in hand analyzes only a very small part of the judgment formation process related to macroeconomic forecasts and there are several areas to be explored, of which some have been mentioned earlier. In this context, the organizational forecast literature is more developed and some of the features that have been evidenced may also be relevant to macroeconomic forecasts and reveal the extent to which they are affected by economic and psychological factors.

# Figures

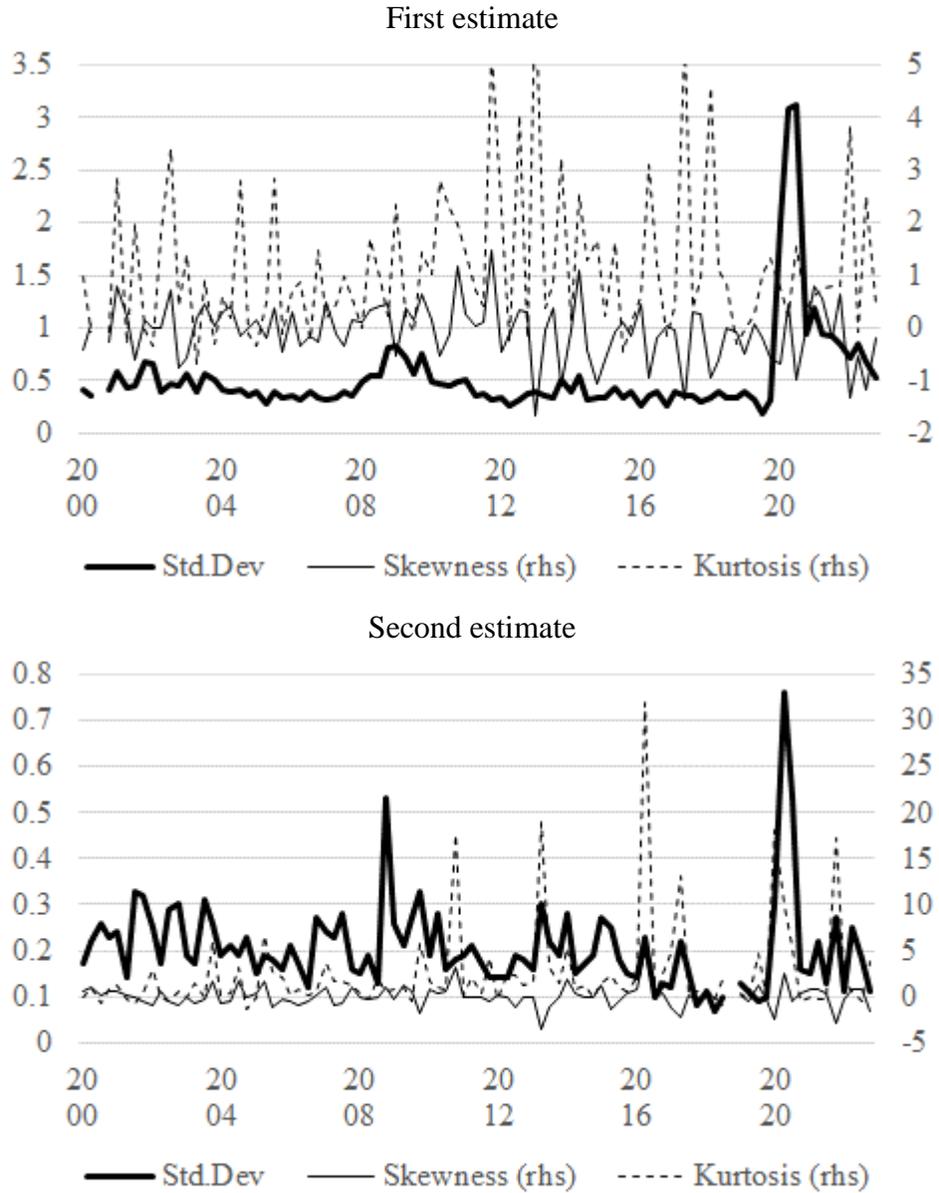

**Figure 1. Moments of the backcast distributions**

Third estimate

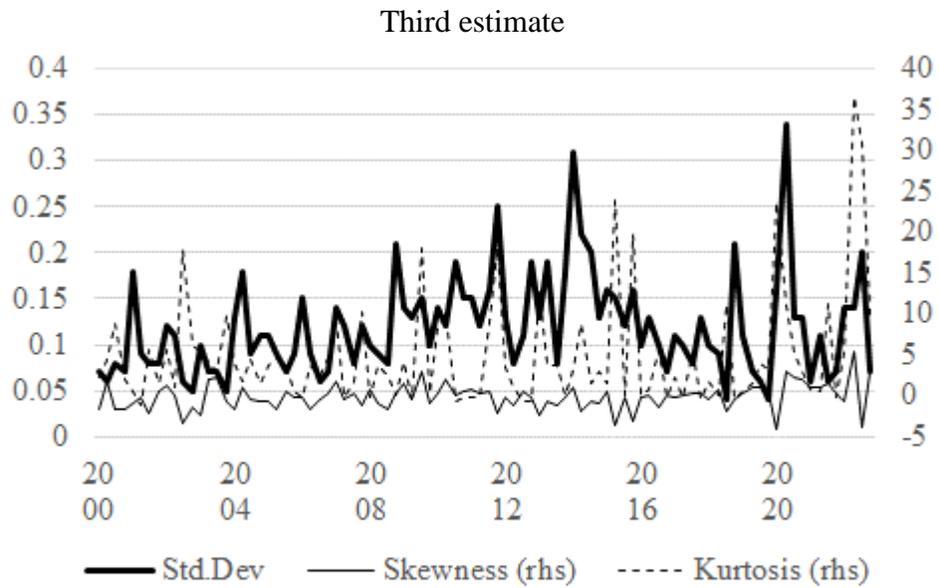

Note: "rhs": Right-hand-side axis.
Source: Own elaboration with observations from Bloomberg.

**Figure 2. Root Mean Squared Errors (RMSE)**

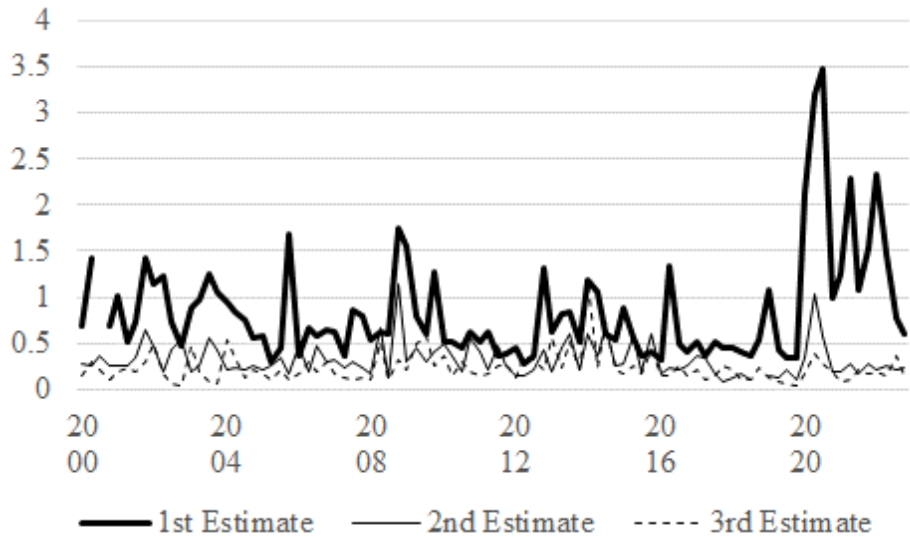

Source: Own elaboration with observations from Bloomberg.

**Figure 3. Distributions of implicit negative judgments**
(percentages of economist with negative judgments)

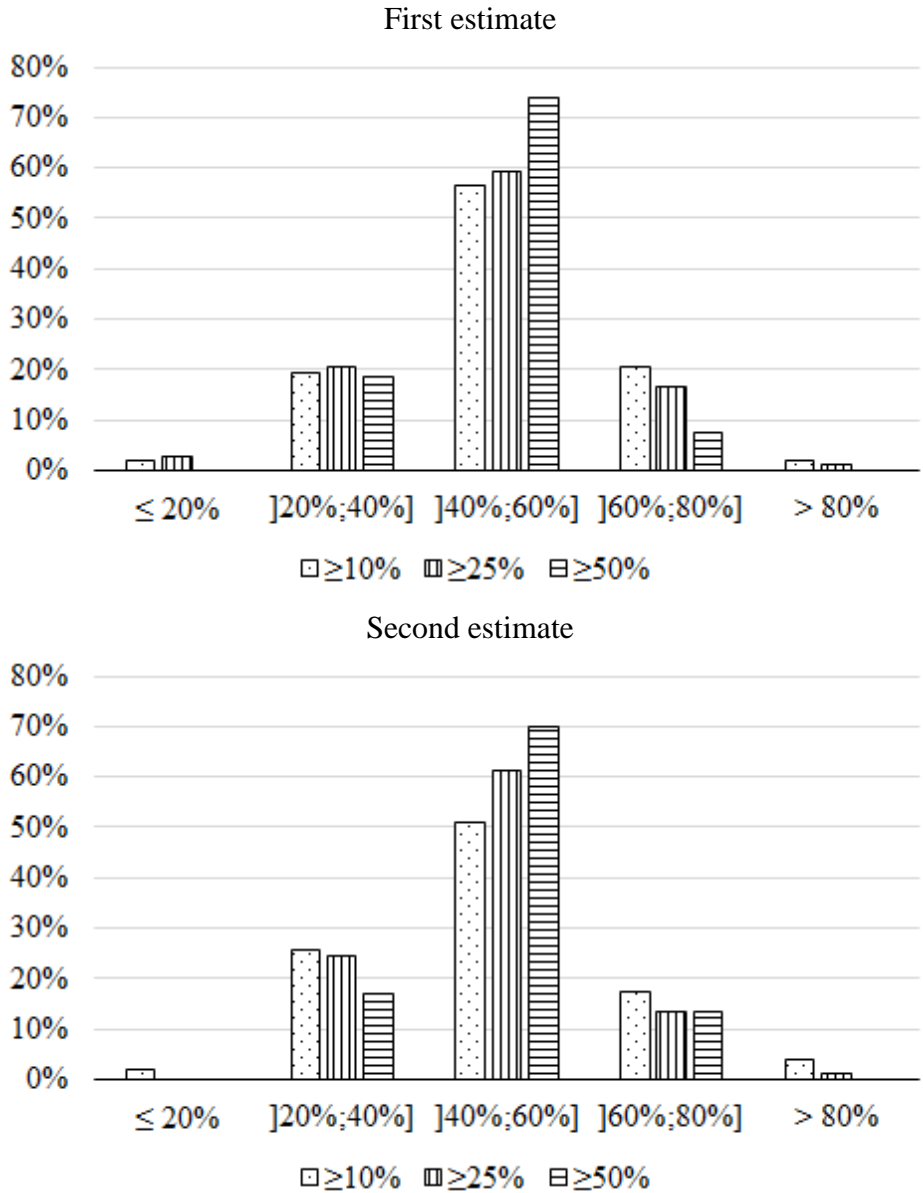

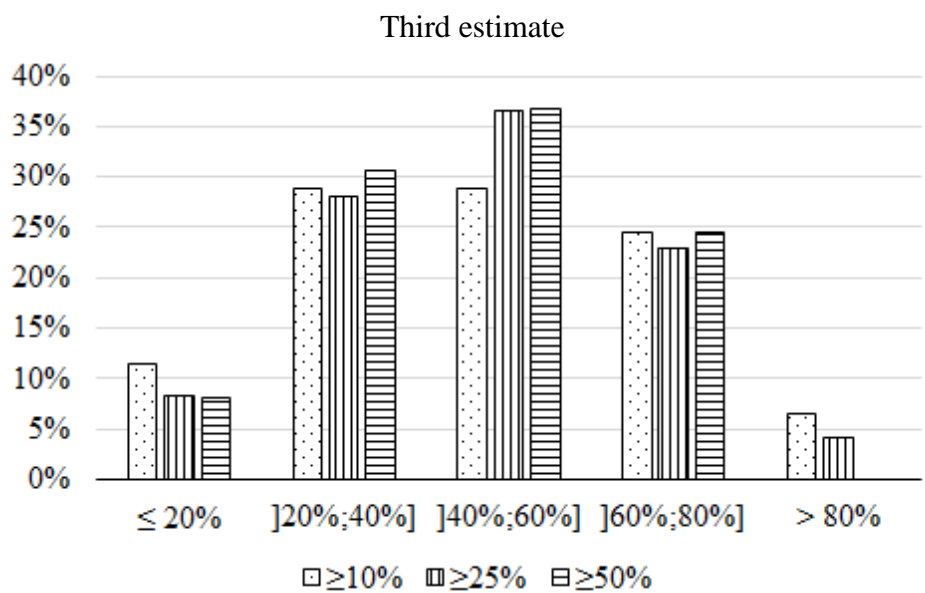

Third estimate

Notes: The figures illustrate the percentages of economists that have applied negative judgment in the relative share of the projections shown on the first axis. For example, the bars related to "≤20" are the shares of analysts whose implicit non-neutral judgment is negative in 20% or less of the reported backasts. The bars are for samples where the forecasters have delivered projections at least 10, 25 and 50% of the quarters.

# Tables

**Table 1. Descriptive statistics of the Bloomberg predictions. 2000Q1-2022Q4**

|  | #Predictions | RMSE | Std.Dev | Skewness | Kurtosis |
|---|---|---|---|---|---|
| 1$^{st}$ estimate | 75.1 | 0.85 | 0.26 | -0.06 | 1.13 |
|  | (37; 87) | (0.28; 3.48) | (0.19; 3.12) | (-1.68; 1.49) | (-0.67; 6.35) |
| 2$^{nd}$ estimate | 69.8 | 0.32 | 0.18 | 0.02 | 2.55 |
|  | (33; 85) | (0.08; 1.13) | (0.07; 0.76) | (-3.46; 4.89) | (-1.30; 32.07) |
| 3$^{rd}$ estimate | 66.9 | 0.23 | 0.08 | -0.15 | 5.02 |
|  | (33; 83) | (0.04; 1.18) | (0.04; 0.34) | (-4.14; 5.51) | (-1.21; 36.61) |

Notes: The table reports the averages across quarters. The following abbreviations are applied: "# Predictions": Number of predictions; "RMSE": Root Mean Squared Error; "Std.Dev": Standard deviation. The numbers in parentheses are for the quarters with the lowest and highest observations.
Source: Own elaboration with observations from Bloomberg.

**Table 2. Number of predictions by specific economists**

|  | 1st estimate | 2nd estimate | 3rd estimate |
|---|---|---|---|
| Total predictions | 6,307 (272) | 5,935 (254) | 5,742 (255) |
|  | | 5,501 (237) | |
|  | | | 5,195 (233) |
|  | | | 4,906 (220) |
| Number of economists with predictions | | | |
| 50% or more of the quarters | 54 | 53 | 49 |
| 25% or more of the quarters | 103 | 98 | 96 |
| 10% or more of the quarters | 156 | 149 | 141 |

Note: Numbers in parentheses are the quantity of economists who made the predictions.
Source: Own elaboration with observations from Bloomberg.

**Table 3. Signs of implicit judgments**
(average percentages across individual forecasters)

|  | Negative | | | Positive | | | Neutral | | |
|---|---|---|---|---|---|---|---|---|---|
|  | ≥10% | ≥25% | ≥50% | ≥10% | ≥25% | ≥50% | ≥10% | ≥25% | ≥50% |
| 1st estimate | 44% | 43% | 43% | 42% | 43% | 43% | 13% | 14% | 14% |
|  | (13%) | (11%) | (9%) | (12%) | (11%) | (8%) | (8%) | (6%) | (5%) |
| 2nd estimate | 36% | 35% | 35% | 35% | 37% | 35% | 29% | 29% | 29% |
|  | (12%) | (10%) | (9%) | (12%) | (10%) | (8%) | (12%) | (11%) | (10%) |
| 3rd estimate | 21% | 23% | 23% | 24% | 25% | 27% | 55% | 52% | 50% |
|  | (11%) | (10%) | (9%) | (15%) | (14%) | (14%) | (19%) | (17%) | (18%) |

Notes: The table reports average percentages of times the implicit judgments of the individual forecasters' predictions are negative, positive or neutral, when employing the median projection as the baseline for all of them. Numbers in parentheses are standard errors. The second row refers to how many times the respondents have supplied predictions during the period considered, at least 10, 25 and 50%, respectively.

**Table 4. Projection efficiency and accuracy**

|  | 1st estimate | | 2nd estimate | | 3rd estimate | |
|---|---|---|---|---|---|---|
|  | Median | Mean | Median | Mean | Median | Mean |
| Unbiasedness | 0.53 | 0.61 | 0.31 | 0.32 | 0.99 | 0.98 |
| Efficiency | 0.00 | 0.00 | 0.01 | 0.00 | 0.00 | 0.02 |
| RMSE | 0.586 | 0.561 | 0.078 | 0.083 | 0.069 | 0.065 |

Notes: The table reports in rows three and four the *p*-values for the heteroscedasticity- and autocorrelation-consistent *F*-tests of unbiasedness and efficiency employing the regression (3) and testing the null hypotheses of $\alpha^k = \beta^k = 0$ and ($\alpha^k = \beta^k = 0$ and $\gamma^k = 0$), respectively. The variable $W_t^k$ includes the most recent median or mean nowcasts from the Survey of Professional Forecasters (SPF) and forecasts made with a recursively estimated AR(1) model. The last row details the Root Mean Squared Error (RMSE) of the predictions of the median and the mean, respectively

### Table 5. Individual forecasters with unbiased and efficient predictions

| #Predictions | 1st estimate | | 2nd estimate | | 3rd estimate | |
|---|---|---|---|---|---|---|
| | Unbiased | Efficient | Unbiased | Efficient | Unbiased | Efficient |
| ≥10% | 74% | 47% | 76% | 54% | 88% | 55% |
| ≥25% | 72% | 45% | 52% | 34% | 62% | 38% |
| ≥50% | 74% | 39% | 30% | 19% | 33% | 17% |

Notes: The table reports the percentage of the forecasters for which the hypotheses of unbiased and efficient predictions could not be rejected when applying the regression (3) and heteroskedasticity robust test statistics. The column "#Predictions" indicates how many of the projections the forecasters have made.

### Table 6. Estimation results. Dependent variable: $j_{i,t}^1$

| | (1) | (2) | (3) | (4) | (5) | (6) |
|---|---|---|---|---|---|---|
| $j_{i,t-1}^1$ | 0.12*** | 0.09*** | 0.10*** | | | |
| | (0.03) | (0.03) | (0.03) | | | |
| $j_{i,t-1}^3$ | | | | 0.11 | 0.07 | 0.11 |
| | | | | (0.08) | (0.08) | (0.09) |
| FE | | X | X | | X | X |
| TE | | | X | | | X |
| #obs | | 5,627 | | | 5,320 | |
| #forecasters | | 236 | | | 236 | |
| $R^2$ | 0.015 | 0.015 | 0.031 | 0.001 | 0.001 | 0.019 |

Notes: Numbers in parentheses are heteroscedastic robust standard errors clustered on forecasters. */**/***: $p$ > 10% / 5% / 1%. "FE": Regression includes fixed effects. "TE": Regression includes time fixed effects. "#obs": Number of observations. "#forecasters": Number of forecasters. Fixed-effects estimators are employed in regressions with fixed effects.

### Table 7. Estimation results. Dependent variable: $j_{i,t}^2$

| | (1) | (2) | (3) | (4) | (5) | (6) |
|---|---|---|---|---|---|---|
| $j_{i,t-1}^2$ | 0.04 | -0.02 | -0.02 | | | |
| | (0.03) | (0.03) | (0.03) | | | |
| $j_{i,t}^1$ | | | | 0.04*** | 0.04*** | 0.04*** |
| | | | | (0.01) | (0.01) | (0.01) |
| FE | | X | X | | X | X |
| TE | | | X | | | X |
| #obs | | 5,165 | | | 5,501 | |
| #forecasters | | 221 | | | 237 | |
| $R^2$ | 0.002 | 0.002 | 0.056 | 0.013 | 0.013 | 0.066 |

Notes: Numbers in parentheses are heteroscedastic robust standard errors clustered on forecasters. */**/***: $p$ > 10% / 5% / 1%. "FE": Regression includes fixed effects. "TE": Regression includes time fixed effects. "#obs": Number of observations. "#forecasters": Number of forecasters. Fixed-effects estimators are employed in regressions with fixed effects.

**Table 8. Estimation results. Dependent variable:** $j_{i,t}^3$

|  | (1) | (2) | (3) | (4) | (5) | (6) |
|---|---|---|---|---|---|---|
| $j_{i,t-1}^3$ | 0.05* | -0.00 | -0.02 |  |  |  |
|  | (0.03) | (0.03) | (0.03) |  |  |  |
| $j_{i,t}^2$ |  |  |  | 0.08*** | 0.07*** | 0.06*** |
|  |  |  |  | (0.02) | (0.02) | (0.02) |
| FE |  | X | X |  | X | X |
| TE |  |  | X |  |  | X |
| #obs |  | 4,993 |  |  | 5,195 |  |
| #forecasters |  | 211 |  |  | 233 |  |
| $R^2$ | 0.003 | 0.003 | 0.071 | 0.019 | 0.019 | 0.088 |

Notes: Numbers in parentheses are heteroscedastic robust standard errors clustered on forecasters. */**/***: $p$ > 10% / 5% / 1%. "FE": Regression includes fixed effects. "TE": Regression includes time fixed effects. "#obs": Number of observations. "#forecasters": Number of forecasters. Fixed-effects estimators are employed in regressions with fixed effects.